\documentclass[letter,longauth]{aa} 
\bibpunct{(}{)}{;}{a}{}{,}
\usepackage{graphicx}
\usepackage{txfonts}
\usepackage{hyperref}
\hypersetup{colorlinks=true,allcolors={blue},pdfborder={0 0 0}}

\hypersetup{draft}

\begin{document} 
\title{Orbital and atmospheric characterization of the planet within the gap of the PDS\,70 transition disk\thanks{Based on observations collected at the European Organisation for Astronomical Research in the Southern Hemisphere under ESO programmes 095.C-0298, 097.C-0206, 097.C-1001, 1100.C-0481.}}
   \author{A. M\"{u}ller\inst{1}
    \and M. Keppler\inst{1}
    \and Th. Henning\inst{1}
    \and M. Samland\inst{1}
    \and G. Chauvin\inst{2,3}
    \and H. Beust\inst{2}
    \and A.-L. Maire\inst{1}
	\and K. Molaverdikhani\inst{1}
    \and R. van Boekel\inst{1}
    \and M. Benisty\inst{2,3}
    \and A. Boccaletti\inst{4}
    \and M. Bonnefoy\inst{2}
    \and F. Cantalloube\inst{1}
	\and B. Charnay\inst{4}
    \and J.-L. Baudino\inst{5}
    \and M. Gennaro\inst{6}
    \and Z.~C. Long\inst{6}
    \and A. Cheetham\inst{7}
    \and S. Desidera\inst{8}
    \and M. Feldt\inst{1}
    \and T. Fusco\inst{9,10}
    \and J. Girard\inst{2,6}
    \and R. Gratton\inst{8}
    \and J. Hagelberg\inst{11}
    \and M. Janson\inst{1,12}
    \and A.-M. Lagrange\inst{2}
	\and M. Langlois\inst{13,14}
    \and C. Lazzoni\inst{8}
    \and R. Ligi\inst{15}
    \and F. M\'enard\inst{2}
    \and D. Mesa\inst{8,16}
    \and M. Meyer\inst{11,17}
    \and P. Molli\`{e}re\inst{18}
    \and C. Mordasini\inst{19}
    \and T. Moulin\inst{2}
    \and A. Pavlov\inst{1}
    \and N. Pawellek\inst{1,20}
    \and S. P. Quanz\inst{11}
    \and J. Ramos\inst{1}
    \and D. Rouan\inst{4}
    \and E. Sissa\inst{8}
    \and E. Stadler\inst{2}
    \and A. Vigan\inst{21}
    \and Z. Wahhaj\inst{22}
    \and L. Weber\inst{7}
    \and A. Zurlo\inst{23,24,10}
          }
   \institute{Max Planck Institute for Astronomy,
              K\"{o}nigstuhl 17, D-69117 Heidelberg\\
              \email{amueller@mpia.de}
 	      \and Univ. Grenoble Alpes, CNRS, IPAG, F-38000 Grenoble, France
 	      \and Unidad Mixta Internacional Franco-Chilena de Astronom\'{\i}a, CNRS/INSU UMI 3386 and Departamento de Astronom\'{\i}a, Universidad de Chile, Casilla 36-D, Santiago, Chile
 	      \and LESIA, Observatoire de Paris, Universit\'{e} PSL, CNRS, Sorbonne Universit\'{e}, Universit\'{e} Paris Diderot, Sorbonne Paris Cit\'{e}, 5 place Jules Janssen, F-92195 Meudon, France
 	      \and Department of Physics, University of Oxford, Oxford, UK
 	      \and Space Telescope Science Institute, 3700 San Martin Dr. Baltimore, MD 21218, USA
		  \and Geneva Observatory, University of Geneva, Chemin des Mailettes 51, 1290 Versoix, Switzerland
 	      \and INAF - Osservatorio Astronomico di Padova, Vicolo della Osservatorio 5, 35122, Padova, Italy
          \and DOTA, ONERA, Université Paris Saclay, F-91123, Palaiseau France
          \and Aix Marseille Universit\'e, CNRS, LAM (Laboratoire d'Astrophysique de Marseille) UMR 7326, 13388 Marseille, France
 	      \and Institute for Particle Physics and Astrophysics, ETH Zurich, Wolfgang-Pauli-Strasse 27, 8093 Zurich, Switzerland
 	      \and Department of Astronomy, Stockholm University, AlbaNova University Center, 106 91 Stockholm, Sweden
 	      \and Aix Marseille Univ, CNRS, LAM, Laboratoire d’Astrophysique de Marseille, Marseille, France
 	      \and CRAL, UMR 5574, CNRS, Universit\'{e} de Lyon, Ecole Normale Sup\'{e}rieure de Lyon, 46 All\'{e}e d’Italie, F-69364 Lyon Cedex 07, France
          \and INAF-Osservatorio Astronomico di Brera, Via E. Bianchi 46, I-23807 Merate, Italy
          \and INCT, Universidad De Atacama, calle Copayapu 485, Copiap\'{o}, Atacama, Chile
          \and Department of Astronomy, University of Michigan, 1085 S. University Ave, Ann Arbor, MI 48109-1107, USA
 	      \and Leiden Observatory, Leiden University, PO Box 9513, 2300 RA Leiden, The Netherlands
          \and Physikalisches Institut, Universit\"at Bern, Gesellschaftsstrasse 6, 3012 Bern, Switzerland
          \and Konkoly Observatory, Research Centre for Astronomy and Earth Sciences, Hungarian Academy of Sciences, P.O. Box 67,\\ H-1525 Budapest, Hungary
          \and Aix Marseille Univ, CNRS, LAM, Laboratoire d’Astrophysique de
Marseille, UMR 7326, 13388, Marseille, France
          \and European Southern Observatory (ESO), Alonso de C\`ordova 3107,
Vitacura, Casilla 19001, Santiago, Chile
 	      \and N\'{u}cleo de Astronom\'{\i}a, Facultad de Ingenier\'{\i}a y Ciencias, Universidad Diego Portales, Av. Ejercito 441, Santiago, Chile
          \and Escuela de Ingenier\'ia Industrial, Facultad de Ingenier\'ia y Ciencias, Universidad Diego Portales, Av. Ejercito 441, Santiago, Chile
             }

   \date{Received --; accepted --}

  \abstract
    {The observation of planets in their formation stage is a crucial, but very challenging step in understanding when, how and where planets form. PDS\,70 is a young pre-main sequence star surrounded by a transition disk, in the gap of which a planetary-mass companion has been discovered recently. This discovery represents the first robust direct detection of such a young planet, possibly still at the stage of formation.}
    {We aim to characterize the orbital and atmospheric properties of PDS\,70\,b, which was first identified on May 2015 in the course of the SHINE survey with SPHERE, the extreme adaptive-optics instrument at the VLT.}
    {We obtained new deep SPHERE/IRDIS imaging and SPHERE/IFS spectroscopic observations of PDS\,70\,b. The astrometric baseline now covers 6 years which allows us to perform an orbital analysis. For the first time, we present spectrophotometry of the young planet which covers almost the entire near-infrared range (0.96 to 3.8\,$\mu$m). We use different atmospheric models covering a large parameter space in temperature, $\log g$, chemical composition, and cloud properties to characterize the properties of the atmosphere of PDS\,70\,b.}
   {PDS\,70\,b is most likely orbiting the star on a circular and disk coplanar orbit at $\sim$22\,au inside the gap of the disk. We find a range of models that can describe the spectrophotometric data reasonably well in the temperature range between 1000--1600\,K and $\log g$ no larger than 3.5\,dex. The planet radius covers a relatively large range between 1.4 and 3.7\,$R_\mathrm{J}$ with the larger radii being higher than expected from planet evolution models for the age of the planet of 5.4\,Myr.}
	{This study provides a comprehensive dataset on the orbital motion of PDS\,70\,b, indicating a circular orbit and a motion coplanar with the disk. The first detailed spectral energy distribution of PDS\,70\,b indicates a temperature typical for young giant planets. The detailed atmospheric analysis indicates that a circumplanetary disk may contribute to the total planet flux.}

   \keywords{Methods: observational -- Techniques: spectroscopic -- Astrometry -- Planets and satellites: atmospheres -- Planets and satellites: individual: PDS70            }
\maketitle

\section{Introduction}
Our knowledge of the formation mechanism and evolution of planets has developed by leaps and bounds since the first detection of an exoplanet by \citet{1995Natur.378..355M} around the main-sequence star 51\,Peg. However, constraining formation time scales, the location of planet formation, and the physical properties of such objects remains a challenge and was so far mostly based on indirect arguments using measured properties of protoplanetary disks. What really is needed is a detection of planets around young stars, still surrounded by a disk.
Modern coronagraphic angular differential imaging surveys such as SHINE (SpHere INfrared survey for Exoplanets, \citealp{2017sf2a.conf..331C}), which are utilizing extreme adaptive optics, provide the necessary spatial resolution and sensitivity to find such young planetary systems.

In \citet{keppler2018} we reported the first bona fide detection of a giant planet inside the gap of the transition disk around the star \object{PDS\,70} together with the characterization of its protoplanetary disk. PDS\,70 is a K7-type 5.4\,Myr young pre-main sequence member of the Upper-Centaurus-Lupus group \citep{2006A&A...458..317R,2016MNRAS.461..794P} at a distance of 113.43$\pm$0.52\,pc \citep{2016AA...595A...1G,2018arXiv180409365G}. Our determination of the stellar parameters are explained in detail in Appendix~\ref{sec:props}. The planet was detected in five epochs with VLT/SPHERE \citep{2008SPIE.7014E..18B}, VLT/NaCo \citep{2003SPIE.4841..944L,2003SPIE.4839..140R}, and Gemini/NICI \citep{2008SPIE.7015E..1VC} covering a wavelength range from $H$ to $L'$-band. In this paper we present new deep $K$-band imaging and first $Y - H$ spectroscopic data with SPHERE with the goal to put constraints on the orbital parameters and atmospheric properties of PDS\,70\,b.

\section{Observations and data reduction}

\subsection{Observations}
We observed PDS\,70 during the SPHERE/SHINE GTO program on the night of February 24th, 2018. The data were taken in the IRDIFS-EXT pupil tracking mode using the N\_ALC\_YJH\_S (185\,mas in diameter) apodized-Lyot coronagraph \citep{2009AA...495..363M,2011ExA....30...39C}. We used the IRDIS \citep{2008SPIE.7014E..3LD} dual-band imaging camera \citep{2010MNRAS.407...71V} with the K$_1$K$_2$ narrow-band filter pair ($\lambda_{K_1}$ = 2.110 $\pm$ 0.102\,$\mu$m, $\lambda_{K_2}$ = 2.251 $\pm$ 0.109\,$\mu$m). A spectrum covering the spectral range from $Y$ to $H$-band (0.96--1.64\,$\mu$m, $R_{\lambda}=30$) was acquired simultaneously with the IFS integral field spectrograph \citep{2008SPIE.7014E..3EC}. We set the integration time for both detectors to 96\,s and acquired a total time on target of almost 2.5\,hours. The total field rotation is 95.7\degr. During the course of observation the average coherence time was 7.7\,ms and a Strehl ratio of 73\% was measured at 1.6\,$\mu$m, providing excellent observing conditions.

\subsection{Data reduction}
\begin{figure}[!ht]
  \centering
  \includegraphics[scale=0.3]{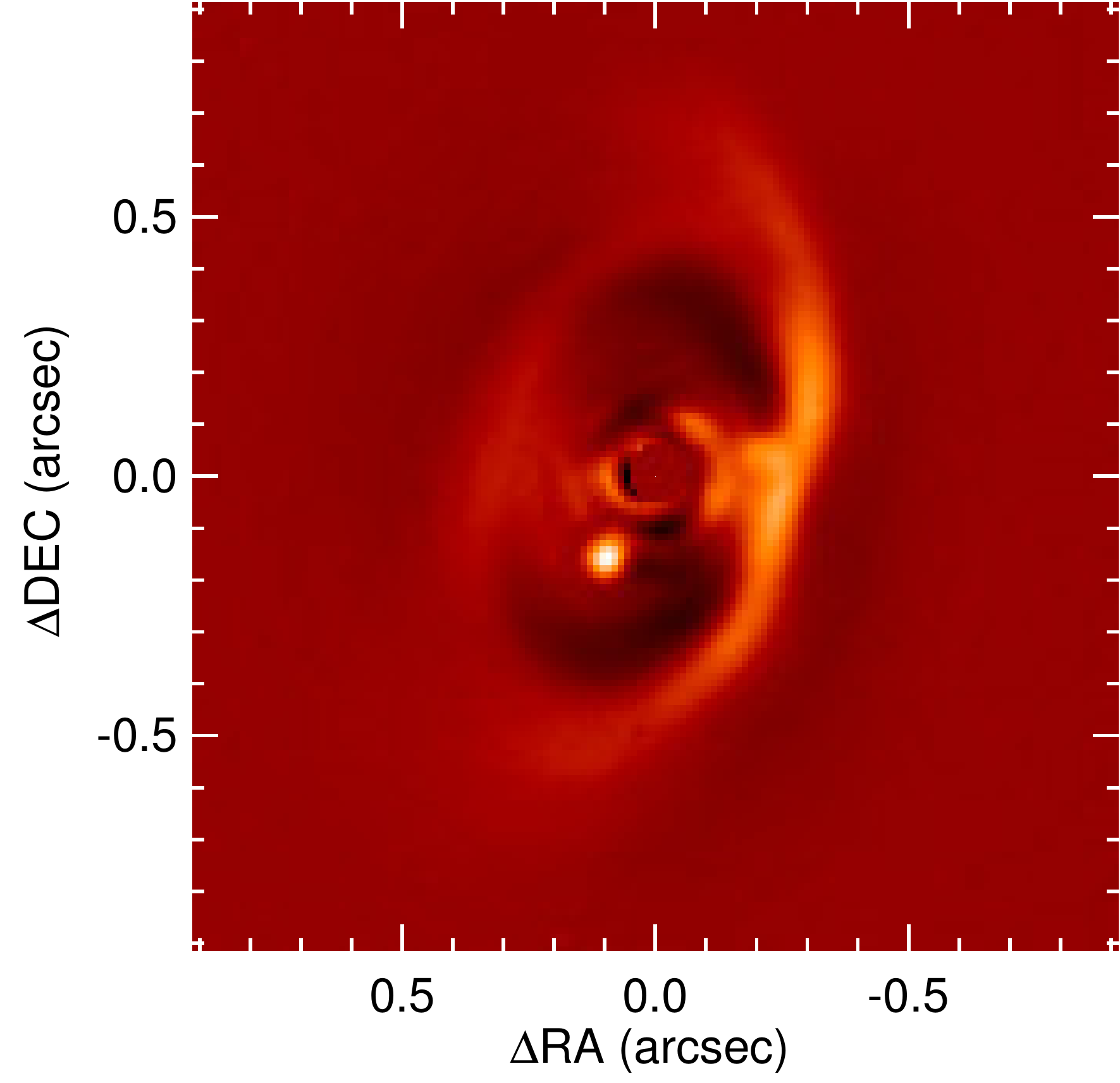}
  \caption{IRDIS combined K$_1$K$_2$ image of PDS\,70 using classical ADI reduction technique showing the planet inside the gap of the disk around PDS\,70. The central part of the image is masked out for better display. North is up, East is to the left. \label{fig:adi}}
\end{figure}

The IRDIS data were reduced as described in \citet{keppler2018}. The basic reduction steps consisted of bad-pixel correction, flat fielding, sky subtraction, distortion correction \citep{2016SPIE.9908E..34M}, and frame registration. \\
The IFS data were reduced with the SPHERE Data Center pipeline {\citep{2017sf2a.conf..347D}}, which uses the Data Reduction and Handling software \citep[v0.15.0,][]{2008SPIE.7019E..39P} and additional IDL routines for the IFS data reduction \citep{2015A&A...576A.121M}. The modeling and subtraction of the stellar speckle pattern for both the IRDIS and IFS data set was performed with an sPCA (smart Principal Component Analysis) algorithm based on \citet{2013A&A...559L..12A} using the same setup as described in \citet{keppler2018}.
Figure~\ref{fig:adi} shows the high-quality IRDIS combined K$_1$K$_2$ image of PDS\,70. The outer disk and the planetary companion inside the gap are clearly visible. In addition, there are several disk related features present, which are further described in Appendix~\ref{app:disk}. For this image the data were processed with a classical ADI reduction technique \citep{2006ApJ...641..556M} to minimize self-subtraction of the disk.
The extraction of astrometric and contrast values was performed by injecting negative point source signals into the raw data (using the unsaturated flux measurements of PDS\,70) which were varied in contrast and position based on a predefined grid created from a first initial estimate of the planets contrast and position. For every parameter combination of the inserted negative planet the data were reduced with the same sPCA setup (maximum of 20 modes, protection angle of 0.75$\times$FWHM) and a $\chi^2$ value within a segment of 2$\times$FWHM and 4$\times$FWHM around the planets position was computed. Following \citet{2016A&A...591A.108O}, the marginalized posterior probability distributions for each parameter was computed to derive final contrast and astrometric values and their corresponding uncertainties (the uncertainties correspond to the 68\% confidence interval). For an independent confirmation of the extracted astrometry and photometry we used SpeCal \citep{2018arXiv180504854G} and find the values in good agreement with each other within 1\,$\sigma$ uncertainty.

\subsection{Conversion of the planet contrasts to physical fluxes}

\begin{figure*}[!ht]
\sidecaption
  \includegraphics[width=12cm]{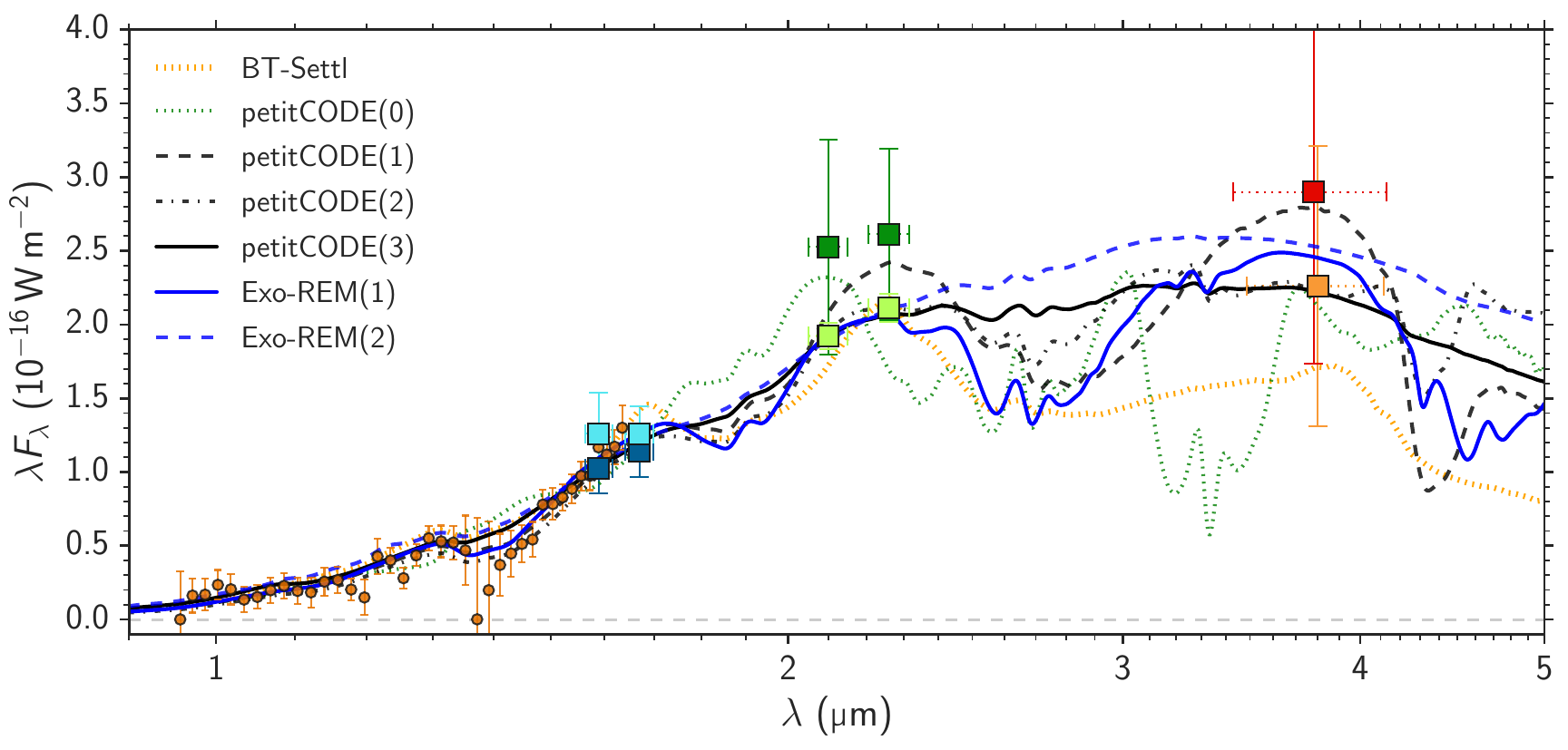}
  \caption{Spectral Energy Distribution of PDS\,70\,b as function of wavelength constructed from $Y$ to $H$-band IFS spectra (orange points), IRDIS H$_2$H$_3$ (first epoch in dark blue, second epoch in light blue) and K$_1$K$_2$ (first epoch in dark green, second epoch in light green), NaCo (red) and NICI (orange) $L'$-band images. Plotted are the best fits for the seven model grids and are smoothed to the resolution of IFS. \label{fig:sed}}
\end{figure*}

The measured contrasts of PDS\,70\,b from all data sets (SPHERE, NaCo, and NICI) were converted to physical fluxes following the approach used in \citet{2016A&A...587A..55V} and \citet{2017A&A...603A..57S}, who used a synthetic spectrum calibrated by the stellar SED to convert the measured planet contrasts at specific wavelengths to physical fluxes. 
In our case, instead of a synthetic spectrum which does not account for any (near-)infrared excess, we made use of the flux calibrated spectrum of PDS\,70 from the SpeX spectrograph \citep{2003PASP..115..362R} which is presented in \citet{2018ApJ...858..112L}. The spectrum covers a wavelength range from 0.7\,$\mu \mathrm{m}$ to 2.5\,$\mu \mathrm{m}$, i.e. the entire IFS and IRDIS data set.
To obtain flux values for our data sets taken in $L'$-band at 3.8\,$\mu \mathrm{m}$ we modeled the stellar SED with simple black bodies to account for the observed infrared excess \citep{2012ApJ...758L..19H,2012ApJ...760..111D}. The final SED of the planet is shown in Fig.~\ref{fig:sed}. The IFS spectrum has a steep slope and displays a few features only, mainly water absorption around $\lambda=1.4\,\mu\mathrm{m}$. The photometric values are listed in Table~\ref{tbl:astphot}.

\section{Results}
\subsection{Atmospheric modeling}
We performed atmospheric simulations for PDS 70\,b with the self-consistent 1D radiative-convective equilibrium tool \textit{petitCODE} \citep{2015ApJ...813...47M,2017A&A...600A..10M}, which resulted in three different grids of self-luminous cloudy planetary model atmospheres (see Table~\ref{tab:model_grids}).
These grids mainly differ in the treatment of clouds: {\it petitCODE}(1) does not consider scattering and includes only Mg$_2$SiO$_4$ cloud opacities, {\it petitCODE}(2) adds scattering, {\it petitCODE}(3) contains four more cloud species including iron (Na$_2$S, KCl, Mg$_2$SiO$_4$, Fe). Additionally, we also use the publicly available cloud-free {\it petitCODE} model grid (here called {\it petitCODE}(0); see \citealp{2017A&A...603A..57S} for a detailed description of this grid) and the public PHOENIX BT-Settl grid \citep{2014IAUS..299..271A,2015A&A...577A..42B}.\\
In order to compare the data to the {\it petitCODE} models we use the same tools as described in \citet{2017A&A...603A..57S}, using the python MCMC code \textit{emcee} on N-dimensional model grids linearly interpolated at each evaluation. We assume a Gaussian likelihood function and take into account the spectral correlation of the IFS spectra \citep{2016ApJ...833..134G}.
For an additional independent confirmation of the results obtained using {\it petitCODE}, we also used cloudy models from the Exo-REM code. The models and corresponding simulations are described in \citet{2018ApJ...854..172C}. Exo-REM assumes non-equilibrium chemistry, and silicate and iron clouds. For the model grid Exo-REM(1) the cloud particles are fixed at 20\,$\mu$m and
the vertical distribution takes into account vertical mixing (with a parametrized $K_{zz}$) and sedimentation.
The model Exo-REM(2) model uses a cloud distribution with a fixed sedimentation parameter $f_{sed}=1$ as in the model by \citet{2001ApJ...556..872A} and {\it petitCODE}.
Table~\ref{tbl:atmosphere} provides a compilation of the best-fit values and Fig.~\ref{fig:sed} shows the respective spectra. The values quoted correspond to the peak of the respective marginalized posterior probability distribution.
The cloud-free models fail to represent the data and result in unphysical parameters. In contrast, the cloudy models provide a much better representation of the data. The results obtained by the {\it petitCODE} and Exo-REM models are consistent with each other. However, because of higher cloud opacities in the Exo-REM(2) models the $\log g$ values are less constrained and the water feature at 1.4\,$\mu$m is less pronounced. Therefore, the resulting spectrum is closer to a black body and the resulting mass is less constrained. All these models indicate a relatively low temperature and surface gravity, but in some cases unrealistically high radii.
Evolutionary models predict radii smaller than 2\,$R_\mathrm{J}$ for planetary-mass objects \citep{2017A&A...608A..72M}.
The radius can be pushed towards lower values, if cloud opacities are removed, e.g. by removing iron ({\it petitCODE}(2)). However, a direct comparison for the same model parameters shows that this effect is very small. In {\it petitCODE}(1) this is shown in an exaggerated way by artificially removing scattering from the models, which leads to a significant reduction in radius. 
In general, we find a wide range of models that are compatible with the current data. The parts of the spectrum most suitable for ruling out models are the possible water absorption feature at 1.4\,$\mu$m, as well as the spectral behavior at longer wavelengths ($K$ to $L'$-band). Given the low signal-to-noise in the water absorption feature and the large uncertainties in the $L'$ flux, it is very challenging to draw detailed physical conclusions about the nature of the object. We emphasize that other possible explanations for the larger than from evolutionary models expected radii, include the recent accretion of material, additional reddening by circumplanetary material, and significant flux contributions from a potential circumplanetary disk. The later possibility is especially interesting in the light of possible features in our reduced images that could present spiral arm structures close to the planet (Fig.~\ref{fig:adi}). There also appears to be an increase in HCO$^{+}$ velocity dispersion close to the location of the planet in the ALMA data presented by \citet{2018ApJ...858..112L}.\\

\begin{table*}
\small
\caption{Model grids used as input for MCMC exploration. The radius of the planet was included as an additional analytic fit-parameter regardless of the model, ranging from $0.1\,R_\mathrm{J}$ to $5 \,R_\mathrm{J}$.}
\centering
\begin{tabular}{l c c c c c c c c p{30mm}}
\hline\hline
Model& $T_\mathrm{eff}$ & $\Delta T$ & $\mathrm{log}\, g$ & $\Delta \mathrm{log}\, g$ &[M/H]& $\Delta \mathrm{[M/H]}$ &$f _\mathrm{sed}$ & $\Delta f _\mathrm{sed} $ & Remarks\\
& $(K)$ & $(K)$ & $\log_{10} \mathrm{(cgs)}$ & $\log_{10} \mathrm{(cgs)}$ & (dex) & (dex) & & &\\
\hline
BT-Settl & 1200 -- 3000 & 100 & 3.0 -- 5.5 & 0.5 & 0.0 & -- & -- & -- & -- \\
{\it petitCODE}(0) & 500 -- 1700 & 50 & 3.0 -- 6.0 & 0.5 & -1.0 -- 1.4 & 0.2 & -- & -- & cloud-free\\
{\it petitCODE}(1) & 1000 -- 1500 & 100 & 2.0 -- 5.0 & 1.0 & -1.0 -- 1.0 & 1.0 & 1.5 & -- & w/o scattering,\newline w/o Fe clouds\\
{\it petitCODE}(2) & 1000 -- 1500 & 100 & 2.0 -- 5.0 & 0.5 & 0.0 -- 1.5 & 0.5 & 0.5 -- 6.0 & 1.0\tablefootmark{a} & with scattering,\newline w/o Fe clouds \\
{\it petitCODE}(3) & 1000 -- 2000 & 200 & 3.5 -- 5.0 & 0.5 & -0.3 -- 0.3 & 0.3 & 1.5 & -- & with scattering,\newline with Fe clouds \\
Exo-REM(1) & 400 -- 2000 & 100 & 3.0 -- 5.0 & 0.1 & 0.32, 1.0, 3.32 & -- & -- & -- & cloud particle\newline size fixed to $20 \muup$m\\
Exo-REM(2) & 400 -- 2000 & 100 & 3.0 -- 5.0 & 0.1 & 0.32, 1.0, 3.32 & -- & 1.0 & -- & --\\
\hline
\end{tabular}
\label{tab:model_grids}
\tablefoot{
\tablefoottext{a}{Except additional grid point at 0.5}
}
\end{table*}

\begin{table*}
  \caption{Parameters of best-fit models based on the grids listed in Table~\ref{tab:model_grids}. The last two columns indicate qualitatively if the corresponding model is compatible with the photometric points in $K$ and $L'$-band, whereas all models describe the $Y$ to $H$-band data well.
  \label{tbl:atmosphere}}
  \centering
  \begin{tabular}{lcccccccc}
  \hline\hline
  Model & $T_\mathrm{eff}$ & log g & [M/H] & $f_\mathrm{sed}$ & Radius & Mass\tablefootmark{b} & $K$ flux & $L'$ flux\\
  & $(K)$ & $\log_{10} \mathrm{(cgs)}$ & (dex) & & $R_\mathrm{J}$ & $M_\mathrm{J}$ & &\\
 \hline
 BT-Settl & 1590 & 3.5 & -- & -- & 1.4 & 2.4 & yes & yes   \\
 {\it petitCODE}(0) & 1155 & 5.5 & -1.0 & -- & 2.7 & 890.0 & no & (yes) \\

 {\it petitCODE}(1) & 1050 & $\leq 2.0$ & $\geq1.0$ & 1.5\tablefootmark{a} & 2.0 & 0.2 & yes & yes  \\
 {\it petitCODE}(2) & 1100 & $2.65$ & $1.0$ & 1.24 & 3.3  & 1.9 & yes & (no)   \\
 {\it petitCODE}(3) & 1190 & $\leq 3.5$ & 0.0 & $\leq 1.5$ & 2.7  & 8.9 & yes & yes   \\
 Exo-REM(1) & 1000 & 3.5 & 1.0 & -- & 3.7 & 17 & yes & yes \\
 Exo-REM(2) & 1100 & 4.1 & 1.0 & 1 & 3.3 & 55 & yes & yes \\
 \hline
\end{tabular}
\tablefoot{
\tablefoottext{a}{Only grid value.}
\tablefoottext{b}{As derived from $\log g$ and radius.}
}
\end{table*}

\subsection{Orbital properties of PDS\,70\,b}

\begin{figure}[!ht]
  \centering
  \includegraphics[scale=0.3]{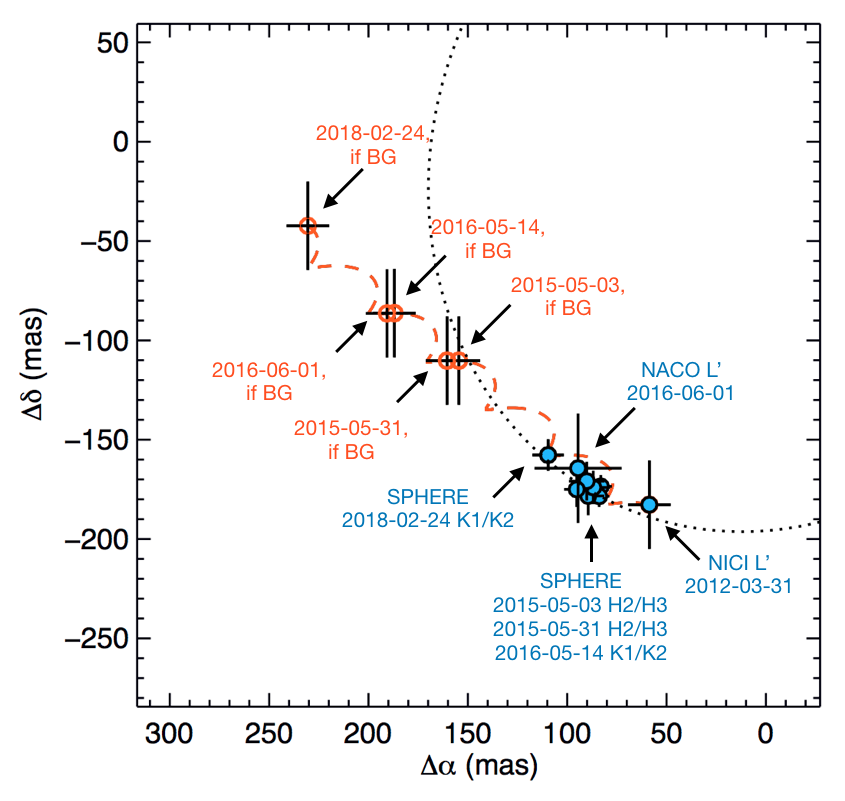}
  \caption{Multi-epoch astrometric measurements of PDS\,70\,b relative to PDS\,70 (marked in blue). The plot shows as well the predictions of the relative position under the hypothesis of a stationary background star for the same observing dates (marked in red). The gray dotted line shows one of the most likely orbital solutions based on our MCMC analysis (see text for details).\label{fig:pm1}}
\end{figure}

The detailed results of the relative astrometry and photometry extracted from our observation from February 2018 are listed in Table~\ref{tbl:astphot} together with the earlier epochs presented in \citet{keppler2018}. A first verification of the relative position of PDS\,70\,b with what we could expect for a stationary background contaminant is shown in Figure~\ref{fig:pm1}. The latest SPHERE observations of February 24th, 2018 confirms that the companion is comoving with the central star.\\
To explore the possible orbital solutions of PDS\,70\,b, we applied the Markov-Chain Monte-Carlo (MCMC) Bayesian analysis technique \citep{2005AJ....129.1706F,2006ApJ...642..505F} developed for $\beta$ Pictoris \citep{2012A&A...542A..41C}, and which is well suited for observations covering a small part of the whole orbit (for large orbital periods) as in the case of PDS\,70\,b. We did not initially consider any prior information on the inclination or longitude of ascending node to explore the full orbital parameter space of bound orbits. As described in Appendix A of \citet{2012A&A...542A..41C}, we assume the prior distribution $p_{0}(x)$ to be uniform in $x = (\log \mathrm{P}, e, \cos i, \Omega+\omega, \omega-\Omega, t_p)$ and work on a modified
parameter vector \vec{u(x)} to avoid singularities in inclination and eccentricities and improve the convergence of the Markov chains. The results of the MCMC analysis are reported in Fig.~\ref{fig:cornerastro}, together with the results of a classical Least-squared linear method (LSLM) flagged by the red line. It shows the standard statistical distribution matrix of the orbital elements $a$, $e$, $i$, $\Omega$, $\omega$, and $t_p$, where $a$ stands for the semi-major axis, $e$ for the eccentricity, $i$ for the inclination, $\Omega$ the longitude of the ascending node (measured
from North), $\omega$ the argument of periastron and $t_p$ the time for periastron passage.
The results of our MCMC fit (Table~\ref{tbl:orbit}) indicate orbital distributions that peak at $22.2^{+6.2}_{-9.7}$\,au (the uncertainties correspond to the 68\% confidence interval) for the semi-major axis, $151.1^{+14.1}_{-13.6}$\,\degr for the inclination, eccentricities are compatible with low-eccentric solutions as shown by the ($a$,$e$) correlation diagram. $\Omega$ and $\omega$ are poorly constrained as low-eccentric solutions are favored and as pole-on solutions are also likely possible. Time at periastron is poorly constrained. The inclination distribution clearly favors retrograde orbits ($i>90\degr$), which is compatible with the observed clockwise orbital motion resolved with SPHERE, NaCo, and NICI.\\
To consider the disk geometry described by \citet{keppler2018}, we decided to explore the MCMC solutions compatible with a planet-disk coplanar configuration. We restrained the PDS70\,b solution set given by the MCMC to those solutions with orbital plane making a tilt angle less than 5\degr~with respect to the disk midplane described by \citet{keppler2018}, i.e., $i=180\degr-49.8\degr$ and $PA=158.6\degr$. The results are shown in Fig.~\ref{fig:cornerastro2} and Table~\ref{tbl:orbit} together with the relative astrometry of PDS\,70\,b\, reported with 200 orbital solutions randomly drawn from our MCMC distributions in Fig.~\ref{fig:mcmcorbitsc}. Figure~\ref{fig:ireldisk} shows the posterior distribution (out of Fig.~\ref{fig:cornerastro}) of the tilt angle with the disk plane assuming $i_\mathrm{disk}=130.2\degr$ and $PA=158.6\degr$. The distribution peaks around $50\degr$, which remains consistent with a likely coplanar planet-disk configuration (or a moderate tilt angle) given the uncertainties. Given the small fraction of orbit covered by our observations, a broad range of orbital configurations are possible including coplanar solutions that could explain the formation of the broad disk cavity carved by PDS\,70\,b. 

\section{Summary and conclusions}
We presented new deep SPHERE/IRDIS imaging data and, for the first time, SPHERE/IFS spectroscopy of the planetary mass companion orbiting inside the gap of the transition disk around PDS\,70.
With the accurate distance provided by {\it Gaia} DR2 we derived new estimates for the stellar mass (0.76$\pm$0.02\,M$_\odot$) and age (5.4$\pm$1.0\,Myr).\\
Taking into account the data sets presented in \citet{keppler2018} we achieve an orbital coverage of 6\,years. Our MCMC Bayesian analysis favors a circular $\sim22$\,au and a disk coplanar wide orbit, which translates to an orbital period of 118\,yr.\\
The new imaging data show rich details in the structure of the circumstellar disk. Several arcs and potential spirals can be identified (see Fig.~\ref{fig:disk}). How these features are connected to the presence of the planet are beyond the scope of this study.\\
With the new IFS spectroscopic data and photometric measurements from previous IRDIS, NaCo, and NICI observations we were able to construct a SED of the planet covering a wavelength range of 0.96 to 3.8\,$\mu$m. We computed three sets of cloudy model grids with the {\it petitCODE} and two models with Exo-REM with different treatment of clouds. These model grids and the BT-Settl grid were fitted to the planets' SED. The atmospheric analysis clearly demonstrates that cloud-free models do not provide a good fit to the data. In contrast, we find a range of cloudy models that can describe the spectrophotometric data reasonably well and result in a temperature range between 1000--1600\,K and $\log g$ no larger than 3.5\,dex. The radius varies significantly between 1.4 and 3.7\,R$_\mathrm{J}$ based on the model assumptions and is in some cases higher than what we expect from evolutionary models. The planets' mass derived from the best fit values ranges from 2 to 17\,M$_\mathrm{J}$, which is similar to the masses derived from evolutionary models by \citet{keppler2018}.\\
This paper provides the first step into a comprehensive characterization of the orbit and atmospheric parameters of an embedded young planet. Observations with JWST and ALMA will provide additional constrains on the nature of this object, especially on the presence of a circumplanetary disk.

\begin{acknowledgements}

SPHERE is an instrument designed and built by a consortium consisting of IPAG (Grenoble, France), MPIA (Heidelberg, Germany), LAM (Marseille, France), LESIA (Paris, France), Laboratoire Lagrange (Nice, France), INAF–Osservatorio di Padova (Italy), Observatoire de Gen\`{e}ve (Switzerland), ETH Zurich (Switzerland), NOVA (Netherlands), ONERA (France) and ASTRON (Netherlands) in collaboration with ESO. SPHERE was funded by ESO, with additional contributions from CNRS (France), MPIA (Germany), INAF (Italy), FINES (Switzerland) and NOVA (Netherlands).  SPHERE also received funding from the European Commission Sixth and Seventh Framework Programmes as part of the Optical Infrared Coordination Network for Astronomy (OPTICON) under grant number RII3-Ct-2004-001566 for FP6 (2004–2008), grant number 226604 for FP7 (2009–2012) and grant number 312430 for FP7 (2013–2016). We also acknowledge financial support from the Programme National de Plan\'{e}tologie (PNP) and the Programme National de Physique Stellaire (PNPS) of CNRS-INSU in France. This work has also been supported by a grant from the French Labex OSUG@2020 (Investissements d’avenir – ANR10 LABX56). The project is supported by CNRS, by the Agence Nationale de la Recherche (ANR-14-CE33-0018). It has also been carried out within the frame of the National Centre for Competence in Research PlanetS supported by the Swiss National Science Foundation (SNSF). MRM, HMS, and SD are pleased to acknowledge this financial support of the SNSF. Finally, this work has made use of the the SPHERE Data Centre, jointly operated by OSUG/IPAG (Grenoble), PYTHEAS/LAM/CESAM (Marseille), OCA/Lagrange (Nice) and Observtoire de Paris/LESIA (Paris). We thank P. Delorme and E. Lagadec (SPHERE Data Centre) for their efficient help during the data reduction process.
This work has made use of the SPHERE Data Centre, jointly operated by OSUG/IPAG (Grenoble), PYTHEAS/LAM/CeSAM (Marseille), OCA/Lagrange (Nice) and Observatoire
de Paris/LESIA (Paris) and supported by a grant from Labex OSUG@2020 (Investissements d’avenir – ANR10 LABX56). A.M. acknowledges the support
of the DFG priority program SPP 1992 "Exploring the Diversity of Extrasolar
Planets" (MU 4172/1-1). F.Me. and M.B. acknowledge funding from ANR of France under contract number ANR-16-CE31-0013. D.M. acknowledges support from the ESO-Government of Chile Joint Comittee program 'Direct imaging and characterization of exoplanets'. J.L.B. acknowledges the support of the UK Science and Technology Facilities Council. A.Z. acknowledges support from the CONICYT + PAI/ Convocatoria nacional subvenci\'on a la instalaci\'on en la academia, convocatoria 2017 + Folio PAI77170087. 
We thank the anonymous referee for a constructive report on the manuscript. This work has made use of data from the European Space Agency (ESA) mission
{\it Gaia} (\url{https://www.cosmos.esa.int/gaia}), processed by the {\it Gaia}
Data Processing and Analysis Consortium (DPAC, \url{https://www.cosmos.esa.int/web/gaia/dpac/consortium}). Funding for the DPAC
has been provided by national institutions, in particular the institutions
participating in the {\it Gaia} Multilateral Agreement. This research has made use of NASA’s Astrophysics
Data System Bibliographic Services of the SIMBAD database, operated
at CDS, Strasbourg. Based on observations obtained at the Gemini Observatory (acquired through the Gemini Observatory Archive), which is operated by the Association of Universities for Research in Astronomy, Inc., under a cooperative agreement with the NSF on behalf of the Gemini partnership: the National Science Foundation (United States), the National Research Council (Canada), CONICYT (Chile), Ministerio de Ciencia, Tecnolog\'{i}a e Innovaci\'{o}n Productiva (Argentina), and Minist\'{e}rio da Ci\^{e}ncia, Tecnologia e Inova\c{c}\~{a}o (Brazil). 
\end{acknowledgements}

\bibliographystyle{aa}
\bibliography{refs}

\begin{appendix}

\section{Determination of host star properties}\label{sec:props}

We use a Markov-Chain Monte-Carlo approach to find the posterior distribution for the PDS\,70 host star parameters, adopting the \textit{emcee} code \citep{2013PASP..125..306F}.
The unknown parameters are the stellar mass, age, extinction, and parallax\footnote[1]{The parallax of PDS\,70 is treated as an unknown parameter in our fit to the host star’s properties, together with mass, age and $A_{V}$. However we imposed a parallax prior, using {\it Gaia} DR2, which strongly constrains the allowed distance values. As a result, the best fit distance value reported here from the MCMC posterior draws is identical to the value provided by the {\it Gaia} collaboration.}, and we assume solar metallicity.
The photometric measurements used for the fit as well as the independently determined effective temperature $T_{\mathrm{eff}}$ and radius are listed in Table~\ref{tbl:stellparams}.
We perform a simultaneous fit of all these observables. The uncertainties are treated as Gaussians and we assume no covariance between them.\\
We use a Gaussian prior from {\it Gaia} for the distance and a Gaussian prior with mean 0.01\,mag and sigma 0.07\,mag, truncated at $A_{V}$=0\,mag, for the extinction \citep{2016MNRAS.461..794P}.
Given $A_{V}$, we compute the extinction in all the adopted bands by assuming a \citet{1989ApJ...345..245C} extinction law.
We use a \citet{2003PASP..115..763C} initial mass function (IMF) prior on the mass and a uniform prior on the age.
The stellar models adopted to compute the expected observables, given the fit parameters, are from the MIST project \citep{2011ApJS..192....3P,2013ApJS..208....4P,2015ApJS..220...15P,2016ApJS..222....8D,2016ApJ...823..102C}. These models were extensively tested against young cluster data, as well as against pre-main sequence stars in multiple system, with measured dynamical masses, and compared to other stellar evolutionary models (see \citet{2016ApJ...823..102C} for details).
The result of the fit constrains the age of PDS\,70 to $5.4\pm1.0$\,Myr and its mass to $0.76\pm0.02$\,M$_\odot$. The best fit parameter values are given by the 50\% quantile (the median) and their uncertainties are based on the 16\% and 84\% quantile of the marginalized posterior probability distribution. The stellar parameters are identical to the values used by \citet{keppler2018}.\\
We note that the derived stellar age of PDS\,70 is significantly younger than the median age derived for UCL with 16$\pm$2\,Myr and an age spread of 7\,Myr by \citet{2016MNRAS.461..794P}. For the computation of the median age \citet{2016MNRAS.461..794P} excluded K and M-type stars for the reason of stellar activity which might bias the derived age. When the entire sample of stars is considered a median age of 9$\pm$1\,Myr is derived. The authors provide an age of 8\,Myr for PDS\,70 based on evolutionary models. Furthermore, the kinematic parallax for PDS\,70 therein is larger by $\sim$15\% compared to the new {\it Gaia} parallax. Thus the luminosity on which the age determination is based on is underestimated and, subsequently, the age is overestimated.

\begin{table}
  \caption{Stellar parameters of PDS\,70. \label{tbl:stellparams}}
  \centering
  \begin{tabular}{lccc}
  \hline\hline
  Parameter & Unit & Value & References\\
  \hline
Distance & pc & 113.43$\pm$0.52 & 1\\
$T_\mathrm{eff}$ & K & 3972$\pm$36 & 2\\
Radius & R$_\odot$ & 1.26$\pm$0.15 & computed from 2\\
$B$ & mag & 13.494$\pm$0.146 & 3\\
$V$ & mag & 12.233$\pm$0.123 & 3\\
$g^\prime$ & mag & 12.881$\pm$0.136 & 3\\
$r^\prime$ & mag & 11.696$\pm$0.106 & 3\\
$i^\prime$ & mag & 11.129$\pm$0.079 & 3\\
$J$ & mag & 9.553$\pm$0.024 & 4\\
$H$ & mag & 8.823$\pm$0.040 & 4\\
$K_\mathrm{s}$ & mag & 8.542$\pm$0.023 & 4\\
Age & Myr & 5.4$\pm$1.0 & this work\\
Mass & M$_\odot$ & 0.76$\pm$0.02 & this work\\
$A_{V}$ & mag & $0.05^{+0.05}_{-0.03}$ & this work\\
  \hline
  \end{tabular}
  \tablebib{(1)~\citet{2016AA...595A...1G,2018arXiv180409365G}; (2) \citet{2016MNRAS.461..794P}; (3) \citet{2015AAS...22533616H}; (4) \citet{2003yCat.2246....0C}.}
\end{table}

\section{The disk seen with IRDIS}\label{app:disk}
\begin{figure}[!ht]
  \centering
  \includegraphics[scale=0.3]{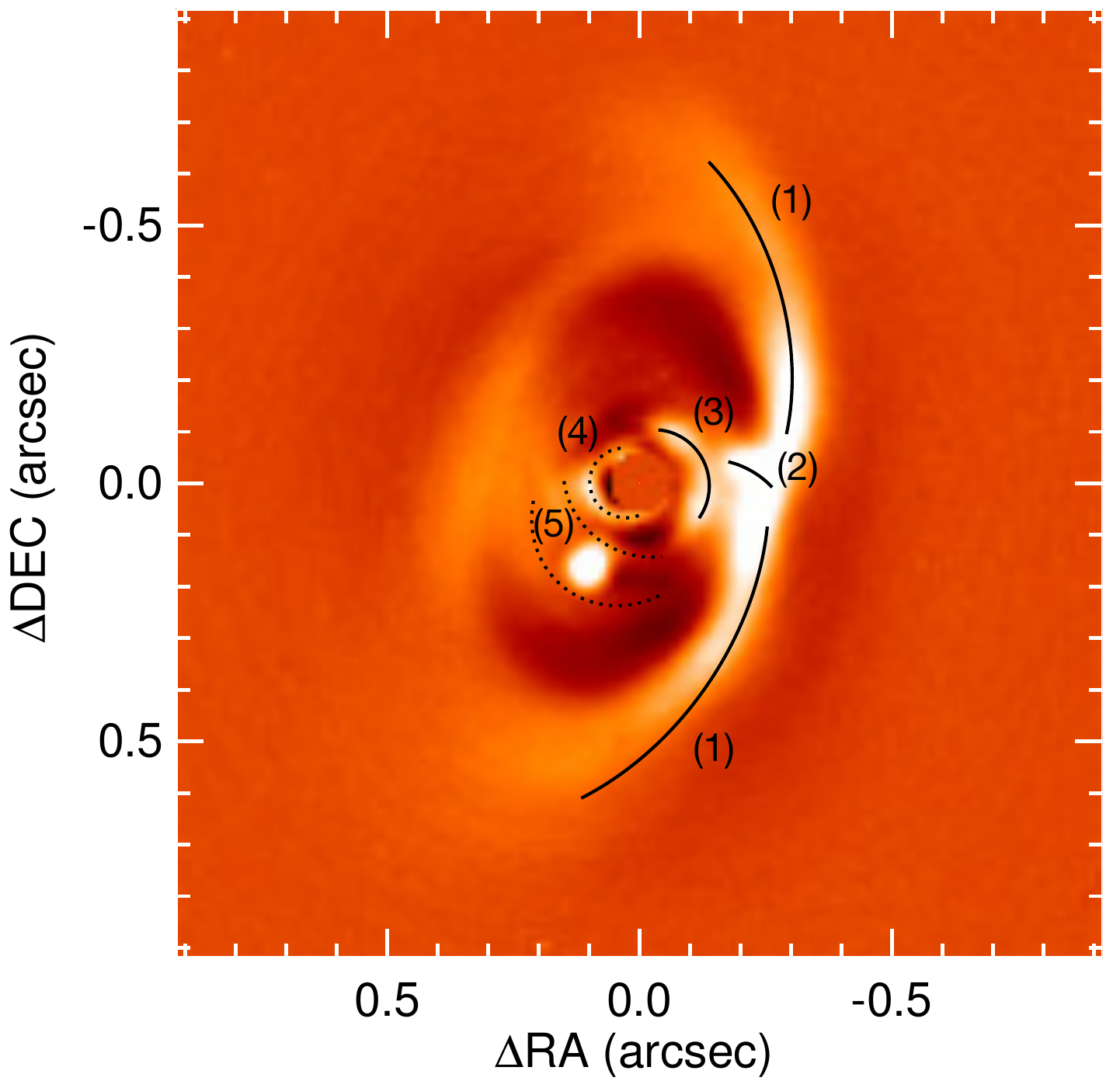}
  \caption{IRDIS combined K$_1$K$_2$ image of PDS\,70 using classical ADI. To increase the dynamic range of the faint disk structures, the companion full intensity range is not shown. The black lines indicate the structures discussed in the above text. North is up, East is to the left. \label{fig:disk}}
\end{figure}

Figure~\ref{fig:disk} shows the IRDIS combined K$_1$K$_2$ image using classical ADI. The image shows the outer disk ring, with a radius of approximately 54\,au, with the West (near) side being brighter than the East (far) side, as in \citet{2012ApJ...758L..19H} and Keppler et al. 2018. The image reveals a highly structured disk with several features: a double ring structure along the West side, which is clearly pronounced along the North-West arc, and which is less but still visible along the South-West side (1), a possible connection from the outer disk to the central region (2), a possible spiral-shaped feature close to the coronagraph (3,4), as well as two arc-like features in the gap on the South East side of the central region (5). Whereas feature (1) and (2) were already tentatively seen in previous observations (see Fig.\,5 and Fig.\,9 in \citealp{keppler2018}), our new and unprecedentedly deep dataset allows one to identify extended structures well within the gap (features 3-5). Future observations at high resolution, i.e. with interferometry will be needed to prove the existence and to investigate the nature of these features, which, if real, would provide an excellent laboratory for probing theoretical predictions of planet-disk interactions. 

\section{Astrometric and photometric detailed results}
\begin{table*}[!ht]
  \caption{Relative astrometry and photometry of PDS\,70\,b as derived from the sPCA reduction. For completeness we list the values from the first five epochs from \citet{keppler2018}. The astrometric values are corrected for True North and accounted for the instrument anamorphism \citep{2016SPIE.9908E..34M}. The True North correction for the IRDIS data recorded on February 24, 2018 is -1.76\degr$\pm$0.06\degr. For True North values from earlier epochs see Table~4 in \citet{keppler2018}. \label{tbl:astphot}}
  \centering
  \begin{tabular}{lccccccccc}
  \hline\hline
  Date & Instr. & Filter & $\Delta\alpha$ (mas) & $\Delta\delta$ (mas) & Sep. [mas] & PA (deg) & $\Delta$mag & mag$_{app}$ & Peak SNR \\
  \hline
2012-03-31 & NICI & L’ & 58.7$\pm$10.7 & -182.7$\pm$22.2 & 191.9$\pm$21.4 & 162.2$\pm$3.7 & 6.59$\pm$0.42 & 14.50$\pm$0.42 & 5.6\\
2015-05-03 & IRDIS & H2 & 83.1$\pm$3.9 & -173.5$\pm$4.3 & 192.3$\pm$4.2 & 154.5$\pm$1.2 & 9.35$\pm$0.18 & 18.17$\pm$0.18 & 6.3\\
2015-05-03 & IRDIS & H3 & 83.9$\pm$3.6 & -178.5$\pm$4.0 & 197.2$\pm$4.0 & 154.9$\pm$1.1 & 9.24$\pm$0.17 & 18.06$\pm$0.17 & 8.1\\
2015-05-31 & IRDIS & H2 & 89.4$\pm$6.0 & -178.3$\pm$7.1 & 199.5$\pm$6.9 & 153.4$\pm$1.8 & 9.12$\pm$0.24 & 17.94$\pm$0.24 & 11.4\\
2015-05-31 & IRDIS & H3 & 86.9$\pm$6.2 & -174.0$\pm$6.4 & 194.5$\pm$6.3 & 153.5$\pm$1.8 & 9.13$\pm$0.16 & 17.95$\pm$0.17 & 6.8\\
2016-05-14 & IRDIS & K1 & 90.2$\pm$7.3 & -170.8$\pm$8.6 & 193.2$\pm$8.3 & 152.2$\pm$2.3 & 7.81$\pm$0.31 & 16.35$\pm$0.31 & 5.5\\
2016-05-14 & IRDIS & K2 & 95.2$\pm$4.8 & -175.0$\pm$7.7 & 199.2$\pm$7.1 & 151.5$\pm$1.6 & 7.67$\pm$0.24 & 16.21$\pm$0.24 & 3.6\\
2016-06-01 & NaCo & L’ & 94.5$\pm$22.0 & -164.4$\pm$27.6 & 189.6$\pm$26.3 & 150.6$\pm$7.1 & 6.84$\pm$0.62 & 14.75$\pm$0.62 & 2.7\\
2018-02-24 & IRDIS & K1 & 109.6$\pm$7.9 & -157.7$\pm$7.9 & 192.1$\pm$7.9 & 147.0$\pm$2.4 & 8.10$\pm$0.05 &  16.65$\pm$0.06 & 16.3\\
2018-02-24 & IRDIS & K2 & 110.0$\pm$7.9 & -157.6$\pm$8.0 & 192.2$\pm$8.0 & 146.8$\pm$2.4 & 7.90$\pm$0.05 & 16.44$\pm$0.05  & 13.7\\
  \hline
  \end{tabular}
\end{table*}

\section{Markov-Chain Monte-Carlo results}\label{app:mcmc}
\begin{figure*}[!ht]
  \centering
  \includegraphics[scale=0.5]{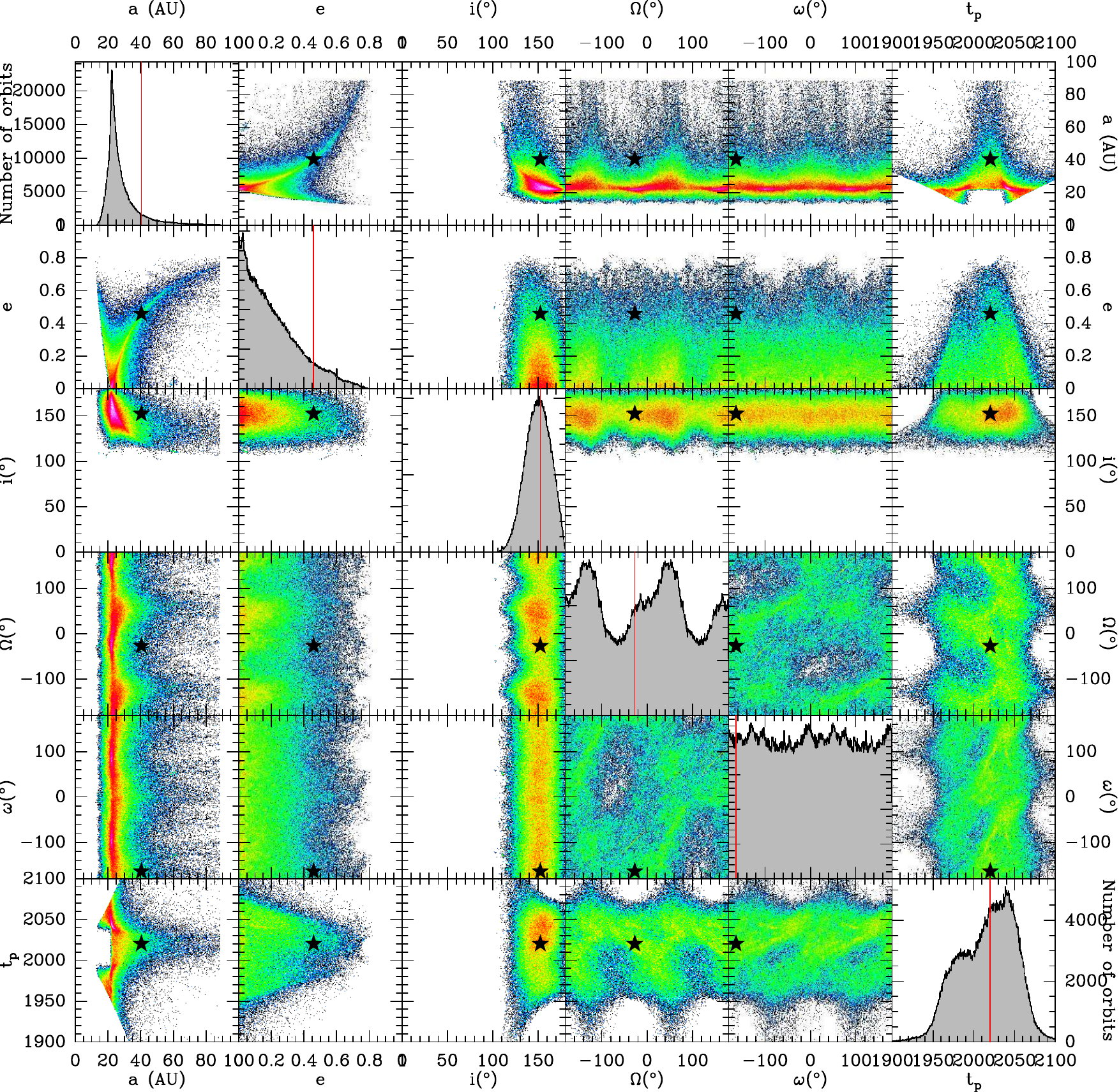}
  \caption{Results of the MCMC fit of the SPHERE, NaCo, and NICI combined astrometric data of PDS\,70\,b reported in terms of statistical distribution matrix of the orbital elements $a$, $e$, $i$, $\Omega$, $\omega$, and $t_P$. The red line in the histograms and the black star in the correlation plots indicate the position of the best LSLM $\chi^{2}_{r}$ model obtained for comparison. \label{fig:cornerastro}}
\end{figure*}

\begin{figure*}[!ht]
  \centering
  \includegraphics[scale=0.5]{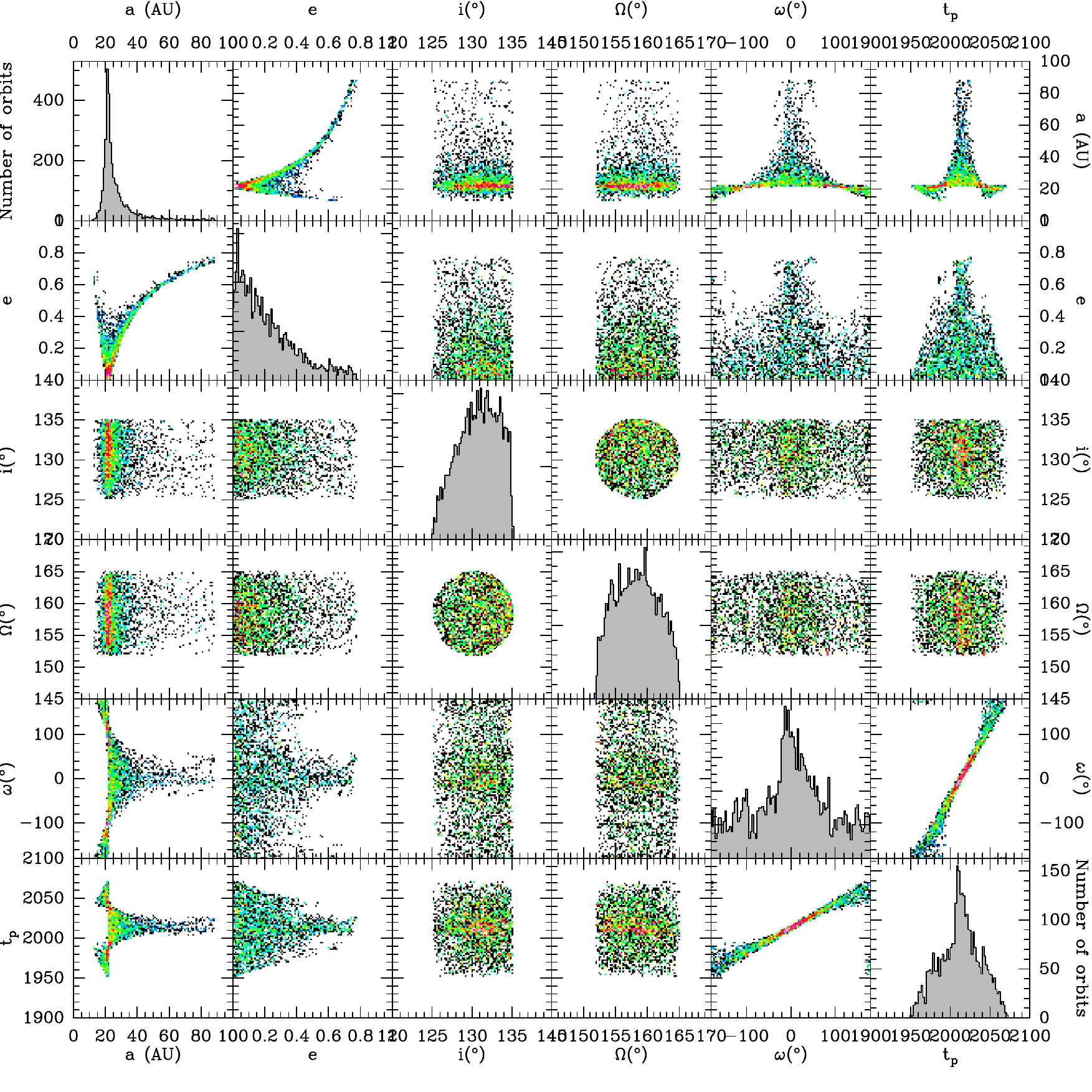}
  \caption{Results of the MCMC fit of the SPHERE, NaCo, and NICI combined astrometric data of PDS\,70\,b reported in terms of statistical distribution matrix of the orbital elements $a$, $e$, $i$, $\Omega$, $\omega$, and $t_P$. We restrained the PDS70\,b solution set given by the MCMC to solutions with orbital plane making a tilt angle less than 5\degr with respect to the disk midplane described by Keppler et al. (2018), i.e., $i=180\degr-49.8\degr$ and $PA=158.6\degr$. 
\label{fig:cornerastro2}}
\end{figure*}

\begin{figure*}[!ht]
  \centering
  \sidecaption
  \includegraphics[width=12cm]{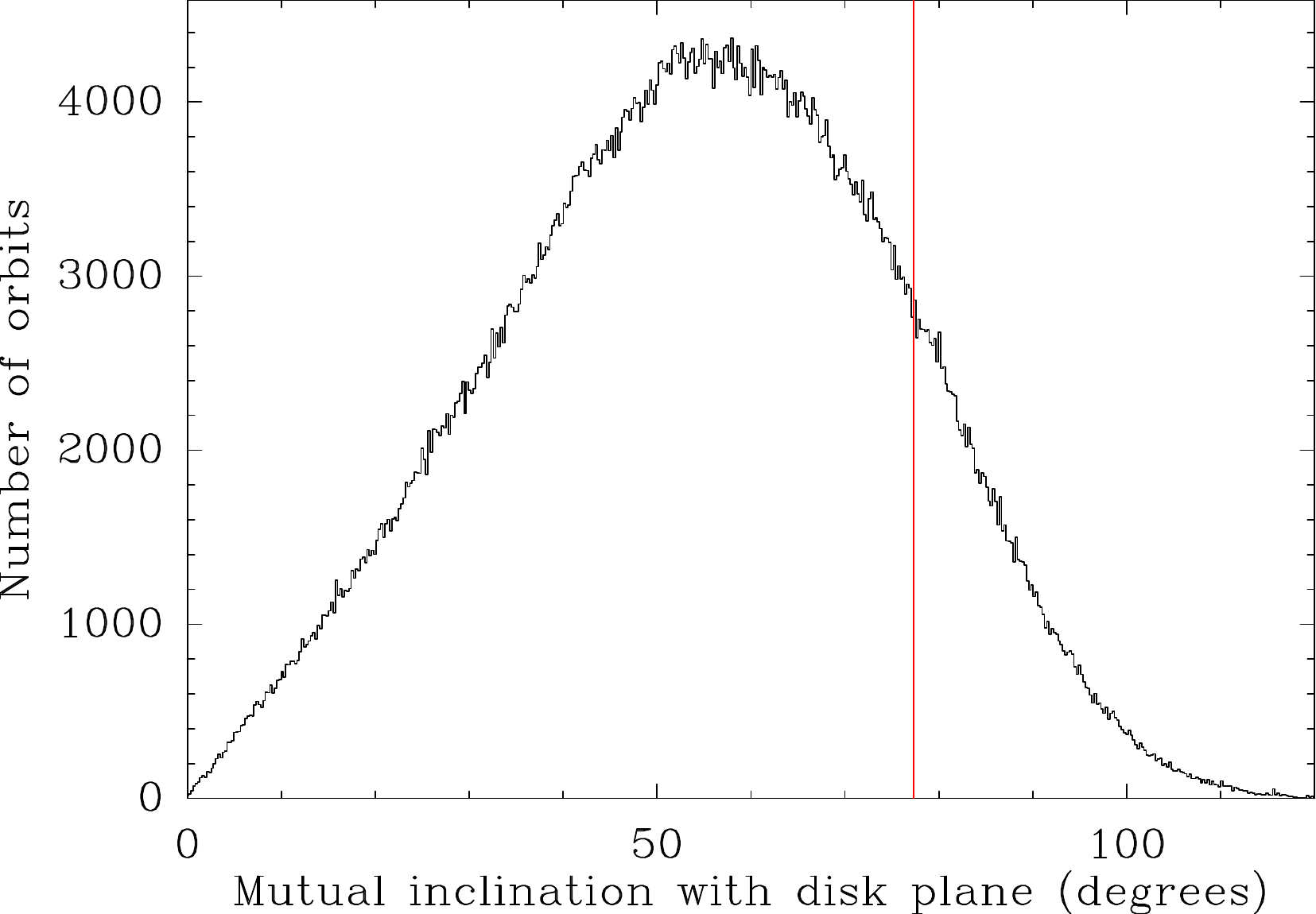}
  \caption{Posterior distribution (out of Fig.~\ref{fig:cornerastro}) of the tilt angle. The distribution peaks around $50\degr$, which remains consistent with a likely coplanar planet-disk configuration. The red line indicates the position of the best LSLM $\chi^{2}_{r}$ model obtained for comparison.
\label{fig:ireldisk}}
\end{figure*}

\begin{table*}[!ht]
  \caption{MCMC solutions for the orbital parameters of PDS\,70\,b. The left part of the table lists the values obtained without any prior information taking into account. The right part of the table lists the solution for the restrained case. The provided lower and upper values correspond to the 68\% confidence interval. \label{tbl:orbit}}
  \centering
  \begin{tabular}{lc|cccc|cccc}
  \hline\hline
  &  & \multicolumn{4}{c|}{unrestrained solutions} & \multicolumn{4}{c}{solutions for restrained $i$ and $\Omega$}\\
  Parameter & Unit & Peak & Median & Lower & Upper & Peak & Median & Lower & Upper \\
\hline
$a$ & au & 22.2 & 25.1 & 12.5 & 28.4 &             21.2 & 23.8 & 13.3 & 27.0 \\
$e$ &  & 0.03 & 0.17 & 0.0 & 0.25 &                 0.03 & 0.18 & 0.0 & 0.27 \\
$i$ & \degr & 151.1 & 150.1 & 137.5 & 165.2 &       131.1 & 131.0 & 128.3 & 133.6 \\
$\Omega$ & \degr & -128.1 & 0.0 & -180.0 & 51.0 &   159.6 & 158.4 & 156.2 & 163.9 \\
$\omega$ & \degr & -130.9 & 0.0 & -180.0 & 59.9 &   -12.7 & 2.5 & -144.7 & 52.3 \\
$t_p$ & yr & 2041.9 & 2020.1 & 2001.4 & 2069.0 &    2009.1 & 2013.4 & 1973.1 & 2029.1 \\
\hline
\end{tabular}
\end{table*}

\begin{figure*}[!ht]
  \centering
  \sidecaption
  \includegraphics[width=12cm]{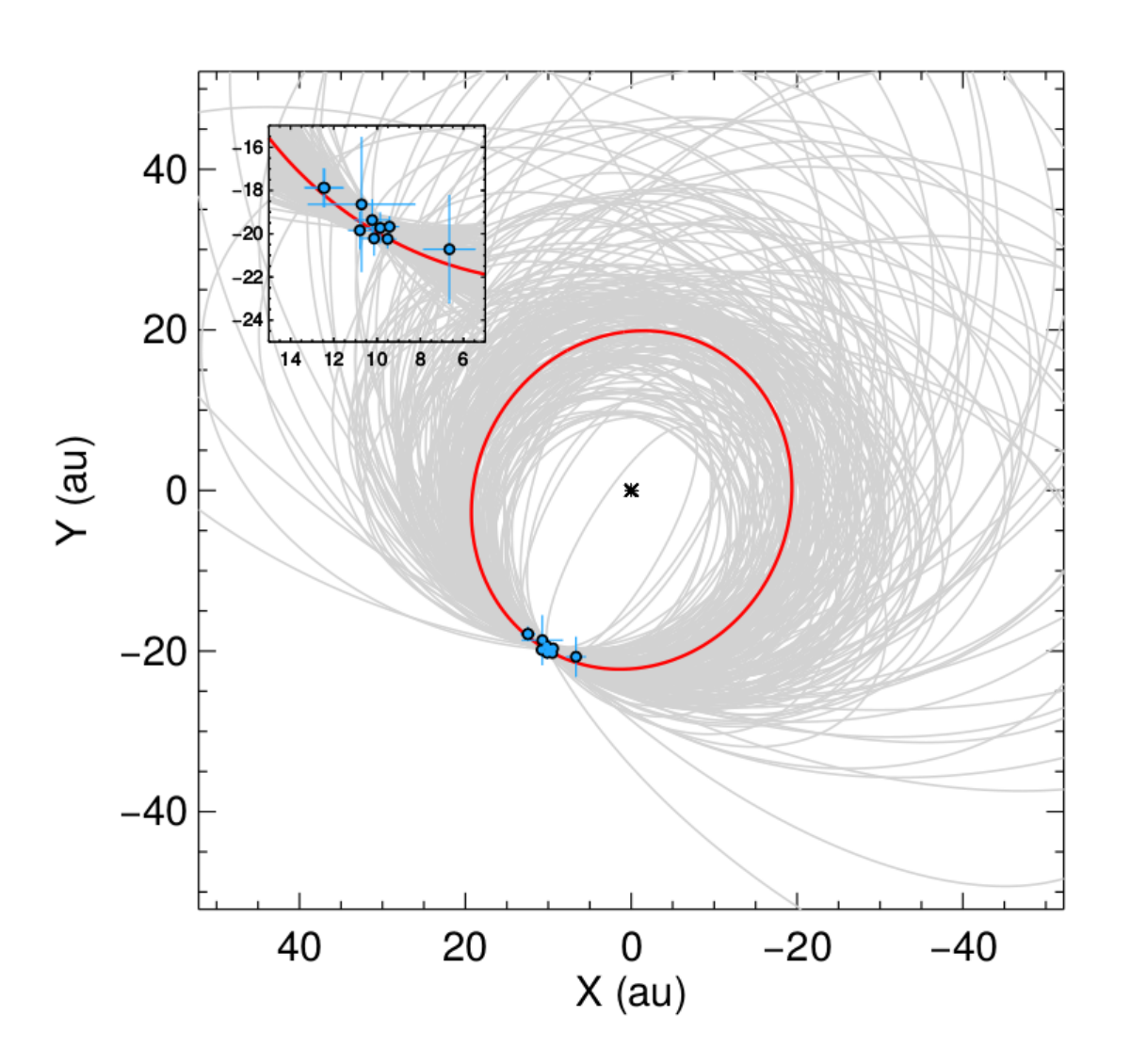}
  \caption{Relative astrometry of PDS\,70\,b reported together with 200 orbital solutions drawn from the MCMC distribution for the coplanar planet-disk configuration. In red is reported one of the most likely solutions from our MCMC analysis as illustration. \label{fig:mcmcorbitsc}}
\end{figure*}

\end{appendix}
\end{document}